# Comments on "Piezonuclear decay of thorium" by F. Cardone et. al.

G. Ericsson[1], S. Pomp[1,*], H. Sjöstrand[1], E. Traneus[2]
[1]Department of Physics and Astronomy, Uppsala University, Box 516, 75121 Uppsala, Sweden
[2]Nucletron Scandinavia, Uppsala, Sweden
[*]Corresponding author at: Department of Physics and Astronomy, Uppsala University, Box 516, 75121 Uppsala, Sweden. E-mail address: Stephan.pomp@physics.uu.se (S.Pomp).

## Abstract

Subjecting a solution of $^{228}$Th to ultrasound (20 kHz, 100W), Cardone et al. [Phys. Lett. A 373 (2009) 1956] claim to observe an increase in the transformation or decay rate of $^{228}$Th by a factor of $10^4$. The evidence provided seems however far from conclusive and in part contradictory to the claims made. In fact, looking at the presented data we find it cannot be taken as justification to discard the null hypothesis, namely, that the data from exposed and non exposed samples are drawn from the same distribution. We suggest a number of additional tests that should be made in order to improve the quality of the study and test the hypothesis of so-called piezonuclear reactions.

## Introduction

In a recent paper, Cardone et al. [1] present a measurement where they subject a solution of $^{228}$Th to cavitation. The authors claim that during the 90-minute experiment, half of the thorium content vanished and conclude that a reduction of $^{228}$Th has occurred that is $10^4$ times faster than the natural decay rate.

In view of the body of knowledge in nuclear physics that has been collected over the past 100 years, this claim is extraordinary and would have rather exciting consequences for the whole field of nuclear physics and its applications. Such an extraordinary claim should, however, be substantiated by extraordinary evidence. We find that such evidence is missing in this paper and it even seems that methodological mistakes have been made.

Since we should, nevertheless, stay open to honest new and potentially revolutionary discoveries, we suggest additional test that the authors can do in order to test their claim.

## Comments regarding the experimental setup and technique

The authors write that the CR39 detectors were "placed underneath the vessel". This means the detectors were placed outside of the borosilicate vessels containing the thorium solution. We note that the range of the emitted α particles in glass is in the order of tens of micrometers and that it thus would be impossible for α particles, and of course even more so for the thorium nuclei themselves, to penetrate the vessel.

Furthermore, the authors claim to recognize thorium alpha decay events by their 'unmistakable "star-shaped" look'. It is not clear to us why decay of $^{228}$Th should be special in this respect; if the star shape indicates detection of several of the 5 alpha decays[1] in the chain to $^{208}$Pb then the same should be true for the 4 alpha decays in the chain of $^{222}$Rn. The authors' reasoning seems based on the failure of the automated counter system "Radosys" to identify the alleged traces as coming from $^{222}$Rn or cosmic rays. To us it seems more logic to conclude that since the events are not recognized by the automated system they are not valid alpha decay events at all. This leads us to wonder how the events were in fact recognized as alpha decay events. If this was done by operator ocular inspection we would strongly caution against operator bias.

To investigate the nature of the observed traces in the CR39 detectors, we suggest that the authors conduct further background measurements, e.g., measurements during cavitation of pure water, or with empty vessels, but in other respects identical to what is done with the thorium solutions. A clearer description of the use of the CR39 detectors and the decay identification procedure would also help.

From the information provided in the paper it seems that each sample solution had a (natural) thorium activity of the order of 2 MBq. This means that for the whole measurement time of 90 minutes and a total of 12 samples about $10^{11}$ of the $^{228}$Th nuclei would decay. The authors claim to have detected and identified 6 such events. It is not clear to us if the authors have made an estimate on whether this is a reasonable amount and if so why this should be the case. Furthermore, if the decay of $^{228}$Th is actually accelerated as (possibly) claimed during the cavitation, and such decay can be registered by the CR39 detectors, then the detectors monitoring the cavitated solutions should not show fewer but four orders of magnitude more events.

Besides investigating the above mentioned questions, we suggest the following measurements to further test the hypothesis. Using their NaI detector, it is a straightforward task to measure the general and characteristic γ radiation that accompanies the decay chain. If the decay rate increases substantially during cavitation, it should be possible to observe an equally increased γ count rate[2]. It is possible, however, that by using the word 'transformation' instead of 'decay', the authors imply that no γ radiation is emitted; that would constitute a second extraordinary claim which would need separate careful study, documentation and verification.

---

[1] An individual $^{228}$Th nucleus decays by a chain of 5 α and 2 β$^-$ decays. There is a branching at the last stages of the chain which means that for an ensemble of $^{228}$Th nuclei, 6 different α decays and 3 different β$^-$ decays will be observed.

[2] A rough estimate shows that an experimenter about one meter away from the setup and assuming a decay rate that is indeed increased by a factor of $10^4$, would be exposed to a dose in the order of hundreds of mSv in the course of the experiment.

Another problem is the lack of reported $^{228}$Th concentration values for each of the test samples, both before and after cavitation. Such measurements are only shown for a selection of 6 samples and only after cavitation.

## Comments regarding the analysis

To draw any positive conclusions from experimental data careful quantitative analyses of the systematic and statistical uncertainties have to be performed and it should be shown that from the data the null hypotheses can be discarded on an acceptable significance level.

In their analysis the authors use Fig. 1 as evidence that the non-cavitated solutions give rise to significantly more events than the cavitated solutions (3 out of 4 plates versus 3 out of 8 plates, respectively). However, the authors have not tested their data against the null hypothesis, namely, that the two data sets are drawn from the same parent distribution. Granted that the authors' identification of thorium decay events is correct, we find that the null hypothesis cannot be rejected on a significance level normally applied in such cases. Performing, for example, a student's t-test we obtain a p-value for the data of 0.26. The p-value is the probability, under the null hypothesis, of observing a value as extreme or more extreme as the test statistic. In order to conform to accepted standards for statistical significance, the null hypothesis can only be rejected if the p-value is very small, say of the order 0.01 or lower; clearly the data presented by the authors fail this test by a large margin. To remedy this shortcoming, we suggest the authors expand their measurements in order to provide much more experimental data and thereby, possibly, substantiate their claim on a level of significance acceptable by the physics community.

In the text the authors state that the "12 identical samples" differed in concentration from 0.01 to 0.03 ppb, i.e., a variation in concentrations of a factor of 3. In Tables 1 and 2 the authors present their results on the concentration of thorium in non-cavitated and cavitated samples, respectively, based on mass spectrometry measurements. Since measured values of the original concentration of each sample are not provided, it is virtually impossible to judge the validity of the treatment of these data. The treatment of the uncertainties in the averaging is also unclear; the authors use an absolute uncertainty of 0.01 ppb both in the individual samples and in the mean. Let us for the sake of argument assume that the quoted uncertainties in the mean value are correctly estimated and also that the original concentration of all samples agree within the quoted uncertainty of 0.01 ppb. If we then treat these uncertainties as 1-sigma uncorrelated uncertainties, the ratio of the two mean values should read R = 2.1 +- 2.6. Thus the uncertainty in this ratio is so large that the null hypothesis, i.e., that the ratio is one, cannot be rejected with any significance.

Furthermore, the paper lacks estimation of the systematical uncertainties and propagation of the uncertainties is not reported. We urge the authors to perform a quantitative analysis on both systematic and statistical uncertainties.

## Conclusions

The authors base their conclusion, that the transformation or decay rate of $^{228}$Th has been increased by a factor $10^4$, on "two converging evidences" regarding the cavitated samples; first, a reduction of alpha traces in the CR39 detectors, and second, a reduced concentration of thorium. However, as we have shown, the authors' own data have by far failed to provide conclusive evidence for such an effect. Due to the mentioned shortcomings in the experimental procedure and the lack of statistical evidence we must dismiss the authors claim as mere speculation.

## References

[1] F. Cardone, R. Mignani, A. Petrucci, Phys. Lett. A 373 (2009) 1956.